\documentclass[12pt]{article}
\begin{document}

\begin{center}
{\huge \bf Renormalizability of the massive Yang-Mills theory} \\[10mm] 
  S.A. Larin \\ [3mm]
 Institute for Nuclear Research of the
 Russian Academy of Sciences,   \\
 60th October Anniversary Prospect 7a,
 Moscow 117312, Russia
\end{center}

\vspace{30mm}

\begin{abstract}
It is shown that the massive Yang-Mills theory is on mass-shell
renormalizable. Thus the Standard Model of electroweak interactions
can be modified by removing terms with the scalar field from the
Lagrangian in the unitary gauge. 
The resulting electroweak theory
without the Higgs particle is on mass-shell renormalizable and unitary.
\end{abstract}

\newpage
The massive Yang-Mills theory \cite{ym} is considered to be
non-renormalizable \cite{boul}, see also \cite{jct,iz} and references
therein. The known way to get renormalizable and  unitary theory
with massive Yang-Mills bosons is due to the Higgs mechanism 
of spontaneous symmetry breaking \cite{pwh}. The mechanism is used
in the Standard $SU(2)\times U(1)$ Model of electroweak interactions
\cite{gws} which is established to be renormalizable \cite{gth}, 
see also \cite{sf} and references therein. 
In this way one introduces in the Model
the scalar Higgs particle which one can hope to see in experiments.

The purpose of the present paper is to show that the massive
Yang-Mills theory is in fact on mass-shell renormalizable.
Hence the Standard Model can be modified by discarding from the
Lagrangian in the unitary gauge all terms containing the scalar field.

Let us consider the massive Yang-Mills theory
of gauge fields $W_{\mu}^a(x)$
defined by the generating
functional of Green functions in the path integral form 
\begin{equation}
\label{odin}
Z(J)=\frac{1}{N}\int dW \exp \left( i \int dx 
\left( -\frac{1}{4}F_{\mu \nu}^a F_{\mu \nu}^a 
+\frac{1}{2} m^2 W_{\mu}^a W_{\mu}^a+J_{\mu}^aW_{\mu}^a \right) \right)
\end{equation}
\[
F_{\mu \nu}^a=\partial_{\mu} W_{\nu}^a -\partial_{\nu}W_{\mu}^a
+g f^{abc}W_{\mu}^b W_{\nu}^c
\]
with the mass $m$ (in general with different masses $m_a$),
$g$ is the coupling constant and $f^{abc}$ are structure
constants of some non-abelian group, e.g. the $SU(n)$ group.

The first term of the Lagrangian is invariant under
the local gauge transformations
\begin{equation}
\label{transfor}
W_{\mu}^a \rightarrow \left(W_{\mu}^a\right)^{\omega}=
 W_{\mu}^a+\partial_{\mu}\omega^a+g f^{abc}W_{\mu}^b \omega^c +O(\omega^2)
\end{equation}
where $\omega^a(x)$ parametrize an element $\Omega(x)$ of the group 
in the usual way $\Omega(x)=exp(i\omega^a T^a)$, $T^a$ being generators of
the corresponding Lie algebra.

The canonical quantization of this theory can cause some problems since
$W_0^a$ are not canonical variables. But the perturbation theory 
can be developed since the field propagator is well defined
\[
D_{\mu \nu}^{ab}=-\frac{\delta^{ab}}{(2\pi)^4}
\frac{g_{\mu \nu}-k_{\mu}k_{\nu}/m^2}{k^2-m^2+i0}
\]
It is the second term of the propagator which does not decrease 
at large $k$ and is assumed to violate renormalizability by
power counting arguments.

To establish on mass-shell renormalizability of the massive Yang-Mills theory
(\ref{odin}) one should show that 
the S-matrix elements
can be made finite by means of counterterms which can be absorbed
into renormalization constants of the parameters $g$ and $m$
although the Green functions are divergent.

We will work within perturbation theory.

To regularize ultraviolet divergences we will use for convenience
dimensional regularization \cite{dr}  with the
space-time dimension $d=4-2\epsilon$, $\epsilon$ being the
regularization parameter.

Let us consider the known model given by the initial
$SU(2)$-invariant Lagrangian possessing the spontaneously broken symmetry
\begin{equation}
\label{spont}
L=-\frac{1}{4}F_{\mu \nu}^a F_{\mu \nu}^a + \left( D_{\mu}\Phi\right) ^+ 
D_{\mu}\Phi -\lambda \left( \Phi^+ \Phi -v^2 \right)^2
\end{equation}
with the doublet of scalar fields $\Phi(x)$ in the fundamental
representation of the group.

Here $D_{\mu} \Phi = \left( \partial_{\mu} -ig \frac{\tau}{2}^a W_{\mu}^a
 \right) \Phi$ is the covariant derivative, 
$\tau^a$ are the Pauli matrices, $\lambda > 0 $,
$v^2 > 0 $.

To get the complete Lagrangian
one makes the shift of the scalar field
\[
 \Phi(x) =\frac{1}{\sqrt{2}} \left( \begin{array}{l}i \phi_{1}(x)
     + \phi_{2}(x)   \\ 
   \sqrt{2}v+ \chi (x) - i \phi_{3}(x) \end{array} \right)
\]
fixes the gauge  and adds ultraviolet counterterms.

Let us consider two gauges: the widely used $R_{\xi}$-gauge
\cite{gth}, \cite{xi} with an arbitrary parameter $\xi$
and the unitary gauge.

In the $R_{\xi}$-gauge one gets the theory described by the generating
functional of Green functions

\[
Z_{R_{\xi}}(J,K)=\frac{1}{N}\int dW d\phi d\chi d\overline{c} dc 
\exp \left( i \int dx \left( L_{R_{\xi}}+J_{\mu}^a
W_{\mu}^a +K \chi \right) \right)
\]
\begin{equation}
\label{rxi}
L_{R_{\xi}}
=-\frac{1}{4}F_{\mu \nu}^a F_{\mu \nu}^a + \frac{m^2}{2} W_{\mu}^a W_{\mu}^a
-mW_{\mu}^a\partial_{\mu}\phi^a
+\frac{1}{2}\partial_{\mu}\phi^a\partial_{\mu}\phi^a
+\frac{1}{2}\partial_{\mu}\chi\partial_{\mu}\chi
\end{equation}
\[
-\frac{M^2}{2}\chi^2
+\frac{g}{2}W_{\mu}^a(\phi^a\partial_{\mu}\chi-\chi\partial_{\mu}\phi^a
+\epsilon^{abc}\phi^b\partial_{\mu}\phi^c)
+\frac{mg}{2}\chi W_{\mu}^a W_{\mu}^a
\]
\[
+\frac{g^2}{8}(\chi^2+\phi^a \phi^a)W_{\mu}^2
-\frac{gM^2}{4m}\chi(\chi^2+\phi^a \phi^a)
-\frac{g^2M^2}{32m^2}(\chi^2+\phi^a \phi^a )^2
\]
\[
-\frac{1}{2\xi}(\partial_{\mu}W_{\mu}^a+\xi m\phi^a)^2
\]
\[
+\partial_{\mu}\overline{c}^a(\partial_{\mu} c^a
-g\epsilon^{abc} c^b W_{\mu}^c)
-\xi m^2\overline{c}^a c^a
-\frac{g}{2}\xi m\chi\overline{c}^a c^a
+\frac{g}{2}\xi m \epsilon^{abc}\overline{c}^a c^b \phi^c 
\]
\[
+counterterms
\]
This theory
describes three physical massive vector bosons  
with the mass $m=gv/\sqrt{2}$, 
and the physical Higgs field $\chi$ with the mass $M=2\lambda v$. 
Here are also Goldstone ghosts $\phi^a$ and Faddeev-Popov ghosts
$c^a$ with masses $\xi m^2$.
The structure of the counterterms  (consistent
with gauge invariance and Slavnov-Taylor identities \cite{s,t} to ensure
unitarity) is well known, see
e.g. \cite{sf}.

This is the renormalizable gauge,
i.e. Green functions are finite.

In the unitary gauge defined by the condition $\phi^a=0$
one has the Lagrangian
\begin{equation}
\label{unitary}
L_U=-\frac{1}{4}F_{\mu \nu}^a F_{\mu \nu}^a 
+ \frac{m^2}{2} W_{\mu}^a W_{\mu}^a
+\frac{1}{2}\partial_{\mu}\chi\partial_{\mu}\chi-\frac{M^2}{2}\chi^2
\end{equation}
\[
+\frac{mg}{2}\chi W_{\mu}^a W_{\mu}^a
+\frac{g^2}{8}\chi^2W_{\mu}^a W_{\mu}^a
-\frac{gM^2}{4m}\chi^3-\frac{g^2M^2}{32m^2}\chi^4
+counterterms
\]
The theory in the unitary gauge is renormalizable only on mass-shell, i.e.
Green functions are divergent at $\epsilon \rightarrow 0$ 
but the S-matrix elements are finite. 
In this gauge all unphysical particles (longitudinal
quanta of vector fields and ghosts) are absent and unitarity of the theory
is manifest.

To show equivalence of S-matrix elements in two gauges one uses
the functional integral technique. Let us repeat it 
for the case of the Landau gauge $\xi=0$ (for simplicity)
which corresponds in fact to the
Lorentz gauge $\partial_{\mu} W_{\mu}^a =0$.
The generating functional of Green functions in the $L$-gauge is
\begin{equation}
\label{zr}
Z_{L}(J,K)=\frac{1}{N}\int dW d\phi d\chi 
\exp \left( i \int dx \left( L_R+J_{\mu}^a
W_{\mu}^a+K\chi \right) \right)\Delta_L(W)\delta(\partial_{\mu} W_{\mu})
\end{equation}
where $\Delta_L(W)$ is the Faddeev-Popov determinant \cite{fp} and $L_R$ is
obtained from $L_{R_{\xi}}$ by omitting terms depending on $\xi$
and $c^a$ (and by corresponding modification
of counterterms). The Lagrangian $L_R$
is invariant under the following gauge transformations
\begin{equation}
W_{\mu}^a \rightarrow \left(W_{\mu}^{\omega}\right)^a=
 W_{\mu}^a+\partial_{\mu}\omega^a+
\tilde{g} f^{abc}W_{\mu}^b \omega^c +O(\omega^2)
\end{equation}
\[
\phi^a \rightarrow (\phi^{\omega})^a=\phi^a -\tilde{m} \omega^a
-\frac{\tilde{g}}{2}f^{abc}\phi^b \omega^c-\frac{\tilde{g}}{2}\chi \omega^a
+O(\omega^2)
\]
\[
\chi \rightarrow \chi^{\omega}=\chi-\frac{\tilde{g}}{2}\phi^a\omega^a
+O(\omega^2)
\]
where 
\[
\tilde{g}=\frac{z_1}{z_2}g~~~ \tilde{m}=\frac{z_1}{z_2}m
\]
and $z_1, z_2$ are renormalization constants of the triple $W$-vertex
and the $W$-field.

One inserts in the functional integral the unity
\begin{equation}
\Delta_U(\chi)\int dw \delta(\phi^{\omega})=1
\end{equation} 
Making the known change of variables
\[
W_{\mu} \rightarrow W_{\mu}^{\omega^{-1}},~~~
\phi \rightarrow \phi^{\omega^{-1}},~~~
\chi \rightarrow \chi^{\omega^{-1}},~~~ \omega^{-1}\rightarrow \omega
\]
and integrating over  $\omega$ one obtains
\begin{equation}
Z_{L}(J,K)=\frac{1}{N}\int dW d\phi d\chi 
\exp \left( i \int dx \left( L_U+J_{\mu}
W_{\mu}^{\tilde{\omega}}
+K\chi^{\tilde{\omega}} \right) \right)\Delta_U(\chi)\delta(\phi)
\end{equation}
where $\tilde{\omega}$ is defined from the equation
\begin{equation}
\label{til}
\partial_{\mu} \left(W_{\mu}^{\tilde{\omega}}\right)^a=
\partial_{\mu}\left( W_{\mu}^a+\partial_{\mu} \tilde{\omega}^a+
\tilde{g} \left(f^{abc}W_{\mu}^b \tilde{\omega}^c\right)\right) +
O(\tilde{\omega}^2)=0
\end{equation}
The Lagrangian $L_U$ is given in eq.(\ref{unitary}). 

The functional $\Delta_U(\chi)$ can be presented on the surface $\phi^a=0$ as
\[
\Delta_U(\chi)= det
\vert\tilde{m}+\frac{\tilde{g}}{2}\chi(x)\vert ^3=  const\cdot exp\left(\int 
\delta^d(0)ln(1+\frac{g}{2m}\chi(x))^3 dx \right)
\] 
In dimensional regularization this functional
is just a constant and can be absorbed in the normalization factor $N$
although this simplification is not essential for the
following derivation.

One obtaines
\begin{equation} 
\label{zunitary}
Z_{L}(J,K)=\frac{1}{N}\int dW d\chi 
\exp \left( i \int dx \left( L_U+J_{\mu}
W_{\mu}^{\tilde{\omega}}
+K\chi^{\tilde{\omega}} \right) \right)
\end{equation}

The expression (\ref{zunitary}) differs from the generating functional
of Green functions in the unitary gauge
\begin{equation} 
Z_{U}(J,K)=\frac{1}{N}\int dW d\chi 
\exp \left( i \int dx \left( L_U+J_{\mu}W_{\mu}
+K\chi \right) \right)
\end{equation}
only by source terms. It is known that this
difference is not essential for the S-matrix elements, see
e.g. \cite{sf}. Thus the physical equivalence of the L-gauge
and the U-gauge is proved.

From eq.(\ref{zunitary}) one sees
that the counterterms of $L_U$ are given by the counterterms of
$L_R$ at $\phi^a(x)=0$.

To consider renormalization for our purpose it is convenient to use
the Bogoliubov-Parasiuk-Hepp subtraction scheme \cite{bph}.
As it is well known in this scheme a counterterm of e.g. a primitively
divergent Feynman diagram is the truncated Taylor expansion of the
diagram itself at some fixed values of external momenta. 
Hence counterterms of mass dependent diagrams are also mass
dependent.

Let us now analyze the dependence of the Green functions on the Higgs
mass $M$. We will use for this purpose the expansion in large $M$.
The algorithm for the large mass expansion of Feynman diagrams
is given in \cite{lrv}, it can be rigorously derived e.g. with the 
technique of \cite{l}. 

The representation (\ref{zunitary}) ensures for the regularized
Green functions of the fields $W$ and $\chi$ that the large
$M$-expansion of $M$-dependent contributions contain either
terms with integer negative 
powers of $M^2$ or terms with non-integer powers of $M^2$
(non-integer powers contain $\epsilon$).
This is because  each vertex  with the
factor $M^2$ has three or four attached $\chi$-lines
due to the structure of $L_U$. Corresponding counterterms (i.e.
counterterms relevant for $L_U$) have the same
property within the large-$M$ expansion.
(In contrast, counterterms of e.g. the four-$\phi$ vertex in $L_R$
contain polynomial in $M$ terms because of the $M^2$-factors in the
couplings of $L_R$.) 

Let us further consider Green functions with external $W$-bosons only.
We will first shortly formulate the result. 
The eq.(\ref{zunitary}) ensures that if one removes
from a renormalized Green function $M$-dependent terms then the remaining
part is finite. On the Lagrangian level it means that one removes
from $L_U$ all terms containing the field $\chi$. Thus 
one obtains the theory 
\begin{equation}
\tilde{Z}(J)=\frac{1}{N}\int dW \exp \left( i \int dx (L_{YM} 
+J_{\mu}W_{\mu}^{\tilde{\omega}}) \right)
\end{equation}
\[
L_{YM}=-\frac{1}{4}F_{\mu\nu}^a F_{\mu\nu}^a
 +m^2 W_{\mu}^a W_{\mu}^a  +counterterms
\]
with finite off-shell Green functions, where $W_{\mu}^{\tilde{\omega}}$
is given by (\ref{til}). 
Since the difference between $W_{\mu}^{\tilde{\omega}}$
and $W_{\mu}$ in the source term is not essential for $S$-matrix
elements the massive Yang-Mills theory is renormalizable on mass-shell.

Let us elaborate these arguments in more detail.
The representation (\ref{zunitary}) ensures, see e.g. \cite{sf},  
that the following on mass-shell expressions for the renormalized Green
functions (relevant for the S-matrix elements) 
should coincide

\begin{equation}
\label{lu}
\left(\frac{1}{\sqrt{z}}\right)^n \prod_{i=1}^{n} (k_i^2-m^2)
G_{\mu_1 ... \nu_n}^{a_1 ... a_n}(k_1 ... k_n)\vert_{L-gauge}
\stackrel{k_i^2=m^2}{=}
\end{equation}
\[
\left(\frac{1}{\sqrt{z}}\right)^n \prod_{i=1}^{n} (k_i^2-m^2)
G_{\mu_1 ... \nu_n}^{a_1 ... a_n}(k_1 ... k_n)\vert_{U-gauge}
\]
where $z$ is the residue of the propagator pole
\begin{equation}
\delta^{ab}\left(g_{\mu \nu}-\frac{k_{\mu}k_{\nu}}{k^2}\right)^2 z=
\lim_{k^2\rightarrow m^2} (k^2-m^2)
\left(g_{\mu \nu}-\frac{k_{\mu}k_{\nu}}{k^2}\right)\int e^{ikx}
\frac{\delta^2 Z(J)}{\delta J_{\mu}^a (x)J_{\nu}^b (0)} dx \vert _{J=0}
\end{equation}

We apply again the large-$M$ expansion to both sides of eq.(\ref{lu})
before removing regularization.
In the L-gauge one can present renormalization in a standard form of
the $R$-operation for individual diagrams. 
This ensures that $M$-dependent terms in the large
$M$-expansion of the l.h.s. of (\ref{lu}) are finite at 
$\epsilon \rightarrow 0$ separately from $M$-independent terms.
Hence in the r.h.s.  $M$-dependent terms are also finite
separately from $M$-independent terms. Thus if one removes all
$M$-dependent terms from the r.h.s. of (\ref{lu}) one is left with a finite
expression.  On the Lagrangian level it means in the unitary gauge
that one removes from $L_U$ all terms containing the field
$\chi$ and also all $M$-dependent terms in the large-$M$ expansion
of counterterms. The resulting theory is on mass-shell finite.
This is the massive Yang-Mills theory
\begin{equation}
\label{dva}
Z(J)=\frac{1}{N}\int dW \exp \left( i \int dx \left(L_{YM} 
+J_{\mu}^aW_{\mu}^a\right) \right)
\end{equation}
\[
L_{YM}=-\frac{1}{4}
z_2(\partial_{\mu} W_{\nu}^a -\partial_{\nu}W_{\mu}^a
+\frac{z_1}{z_2} g f^{abc}W_{\mu}^b W_{\nu}^c)^2
+z_m m^2 W_{\mu}^a W_{\mu}^a
\]

After  renormalizability is established one can fix
renormalization constants $z_1, z_2$ and $z_m$ within the theory
(\ref{dva})  (without referring to the L-gauge) by
proper normalization conditions.

It is known that the Higgs theories of vector mesons posses so
called tree level unitarity, see e.g. \cite{jct} and references
therein. Tree level cross sections of such theories grow at high energies
slowly enough and do not exceed the so called unitary
limit imposed by the unitarity condition. The reversed
statement is also proved:
from the condition of tree level unitarity follows that a theory
of vector mesons should be a Higgs theory \cite{clt}.
But one can see that tree
level unitarity is not the necessary condition for
renormalizability. Tree level unitarity is violated in the massive
Yang-Mills theory. It indicates that higher order contributions
become relevant at high energies.

The above derivation of on mass-shell renormalizability 
is applicable also to other
gauge groups. It can be straightforwardly applied to the
Standard $SU(2)\times U(1)$ Model of electroweak interactions.
The presence of the $U(1)$ gauge boson and of fermions
does not change the derivation. One can remove from the
Lagrangian in the unitary gauge  all terms containing the scalar
field.
The resulting electroweak theory without the Higgs particle is
on mass-shell renormalizable and unitary.\\

The author is grateful to D.S. Gorbunov and S.M. Sibiryakov
for helpful discussions.

\end{document}